\documentclass[conference]{IEEEtran}
\IEEEoverridecommandlockouts

\usepackage{cite}
\usepackage{amsmath,amssymb,amsfonts}
\usepackage{bm}
\usepackage{array,tabularx,multirow}
\usepackage{algorithmic}
\usepackage{graphicx}
\usepackage{textcomp}
\usepackage{booktabs}
\usepackage{amssymb}
\usepackage{amsthm}
\newtheorem{problem}{Problem}
\usepackage[hyphens]{url}
\usepackage{xurl}
\usepackage[table]{xcolor}
\definecolor{rowgray}{gray}{0.93}
\definecolor{ourred}{RGB}{170,30,30}
\definecolor{ourrow}{gray}{0.93}
\def\BibTeX{{\rm B\kern-.05em{\sc i\kern-.025em b}\kern-.08em
    T\kern-.1667em\lower.7ex\hbox{E}\kern-.125emX}}

\begin{document}
\bstctlcite{IEEEcontrol}
\raggedbottom

\title{RayAR: Autoregressive Dominant Path Generation for Low-Overhead MIMO Beam Alignment}

\author{\IEEEauthorblockN{Junxi Huang\IEEEauthorrefmark{1},
Xiucheng Wang\IEEEauthorrefmark{1},
Nan Cheng\IEEEauthorrefmark{1},
Kailong Wang\IEEEauthorrefmark{1},
Jiang Guo\IEEEauthorrefmark{2},
Yuxiang Zhang\IEEEauthorrefmark{3}
}

\IEEEauthorblockA{
\small
\IEEEauthorrefmark{1}
State Key Laboratory of ISN and School of Telecommunications Engineering,
Xidian University, Xi'an 710071, China\\
\IEEEauthorrefmark{2}
School of Physics, Xidian University, Xi'an 710071, China\\
\IEEEauthorrefmark{3}
State Key Laboratory of Networking and Switching Technology,
Beijing University of Posts and Telecommunications, Beijing, China\\
Email: \{jxhuang\_1, xcwang\_1, jiangguo, 24012100069\}@stu.xidian.edu.cn,
dr.nan.cheng@ieee.org, zhangyx@bupt.edu.cn}
}

\maketitle

\begin{abstract}
Beam alignment in multiple-input multiple-output (MIMO) systems requires selecting suitable transmit and receive beam pairs with limited measurement overhead. Although 3GPP Release 18 investigates beam prediction from partial measurements, collecting these measurements still incurs probing overhead. Geometry can focus this probing through the departure and arrival directions of dominant paths, whose identification requires accounting for coupled diffraction interactions. To address this challenge, we propose RayAR, an autoregressive method that generates dominant propagation paths from urban geometry. RayAR conditions candidate ranking on the path prefix, local diffraction geometry, and receiver reachability, updating this context after each selection until the receiver becomes visible. The completed path's departure and arrival directions rank beam pairs for probing, allowing the pair with the highest received power to be selected from a small candidate set. Under 3GPP Release 18 beam-management metrics, RayAR reduces probing by 96.88\% relative to exhaustive sweeping and achieves the highest optimal-pair hit rate among compared methods at the same budget. Real-world measurements further demonstrate the accuracy of this geometric guidance, with simulation-trained RayAR achieving the lowest departure and arrival angular errors among compared methods without fine-tuning and a mean arrival-direction error of 9.07$^\circ$.
\end{abstract}


\section{Introduction}
\label{sec:intro}

\begin{figure}[!t]
\centering
\includegraphics[width=0.95\linewidth,pagebox=cropbox]{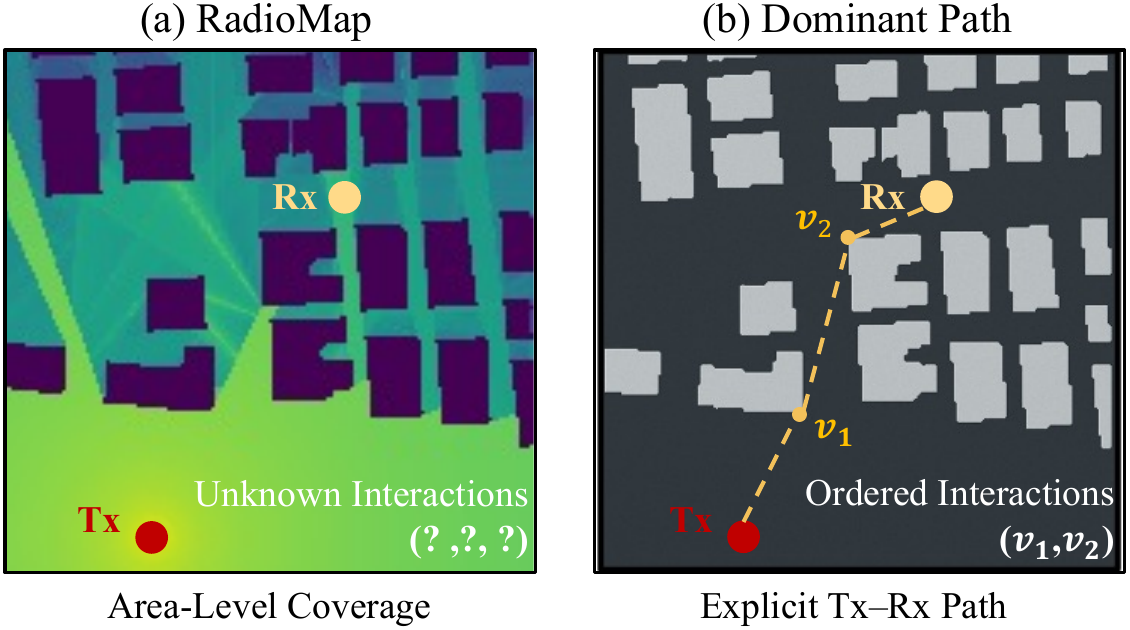}
\caption{From radio maps to explicit propagation paths.}
\label{fig:task}
\end{figure}

Beam alignment in multiple-input multiple-output (MIMO) systems requires selecting beam pairs that deliver high received power while controlling reference-signal and measurement overhead. Exhaustive sweeping establishes a reference optimum by measuring all transmit--receive beam combinations, but its cost grows with the transmit and receive codebooks\cite{heng2022probing}. To reduce this cost, 3GPP Release~18 TR~38.843 studies spatial-domain beam prediction, where measurements from beam Set B guide predictions over beam Set A~\cite{3gpp38843}. Because these predictions require measurements of Set B, the choice of which beams to probe remains important to the overhead--performance tradeoff. When environmental geometry and transmitter--receiver locations are available, a dominant propagation path provides a departure direction (DoD) and a corresponding arrival direction (DoA) for prioritizing beam pairs before measurement, as illustrated in Fig.~\ref{fig:task}. This geometric guidance can concentrate probing on promising candidates, provided that the generated directions accurately reflect real propagation.

An explicit path supports such guidance by connecting the endpoint directions to the interactions that determine propagation loss. Whereas radio maps describe regional path loss~\cite{wang2025radiodiff} and angular power spectra characterize directional power~\cite{huang2026map2aps}, explicit paths additionally identify the intervening interaction sequence and support loss and delay evaluation. Ray tracing~ \cite{kim2024raytracing} and dominant path modeling~\cite{wahl2005dominant} recover this sequence through geometric search and propagation calculations, with learning-based visibility construction~\cite{wang2026d2los} and sequential candidate generation~\cite{jin2025sandwich,eertmans2025sampling} improving search efficiency. For direct dominant path generation, the central decision is which feasible interaction sequence carries strong received power. On urban non-line-of-sight (NLoS) links, routes through different building corners can incur different diffraction losses, so a shorter route can lose its distance advantage through unfavorable interactions~\cite{kouyoumjian1974utd,wahl2005dominant}. Selecting promising beam directions from geometry therefore requires considering how successive interactions jointly determine path strength.

This dependence makes the selected path prefix informative for each subsequent vertex choice. At a building corner, the preceding segment fixes the incoming direction, while the next vertex determines both the outgoing direction and the incoming geometry at the following interaction. To account for this coupling during generation, we propose RayAR, an autoregressive method that generates dominant paths for low-overhead MIMO beam alignment. RayAR reuses a vertex visibility graph across queries and combines the selected prefix with local diffraction geometry and receiver reachability through history-conditioned feature fusion. The resulting candidate representations allow attention to compare legal continuations using the propagation geometry established by earlier choices. Each selection updates that geometry for subsequent decisions until the path reaches a point with visibility to the receiver. Its DoD and DoA then prioritize beam-pair probes, from which measured received power determines the final selection. Because this selection depends on the generated directions, we examine both the resulting probing efficiency and the angular accuracy of simulation-trained RayAR in outdoor measurements. The main contributions of this paper are summarized as follows.
\begin{enumerate}
\interlinepenalty=10000
\item We propose RayAR for low-overhead MIMO beam alignment through on-demand dominant path generation. A reusable visibility graph supports direct generation of an explicit interaction sequence, whose direction of departure (DoD) and direction of arrival (DoA) jointly prioritize transmit and receive beams before probing.
\item We design a history-conditioned autoregressive decoder that allows preceding geometric decisions to shape subsequent candidate ranking. Adaptive feature fusion combines the selected path prefix with local diffraction geometry and receiver reachability, enabling candidate attention to compare continuations under visibility constraints.
\item We demonstrate the effectiveness of RayAR through path estimation, beam-alignment evaluation using 3GPP Release~18 metrics~\cite{3gpp38843}, and outdoor measurements. On RayVerse-100, RayAR achieves 79.69\% exact path-sequence accuracy, and four beam probes yield a higher optimal-pair hit rate and higher mean received power than AR-GAT using eight probes. Across 62 outdoor measurement points covering line-of-sight and NLoS links, simulation-trained RayAR achieves mean arrival and departure angular errors of 9.07$^\circ$ and 11.16$^\circ$, respectively, both the lowest among the compared methods, without fine-tuning. It also generates complete paths satisfying the visibility constraints for all measured NLoS links.
\end{enumerate}

\begin{figure}[!t]
\centering
\includegraphics[width=\linewidth,pagebox=cropbox]{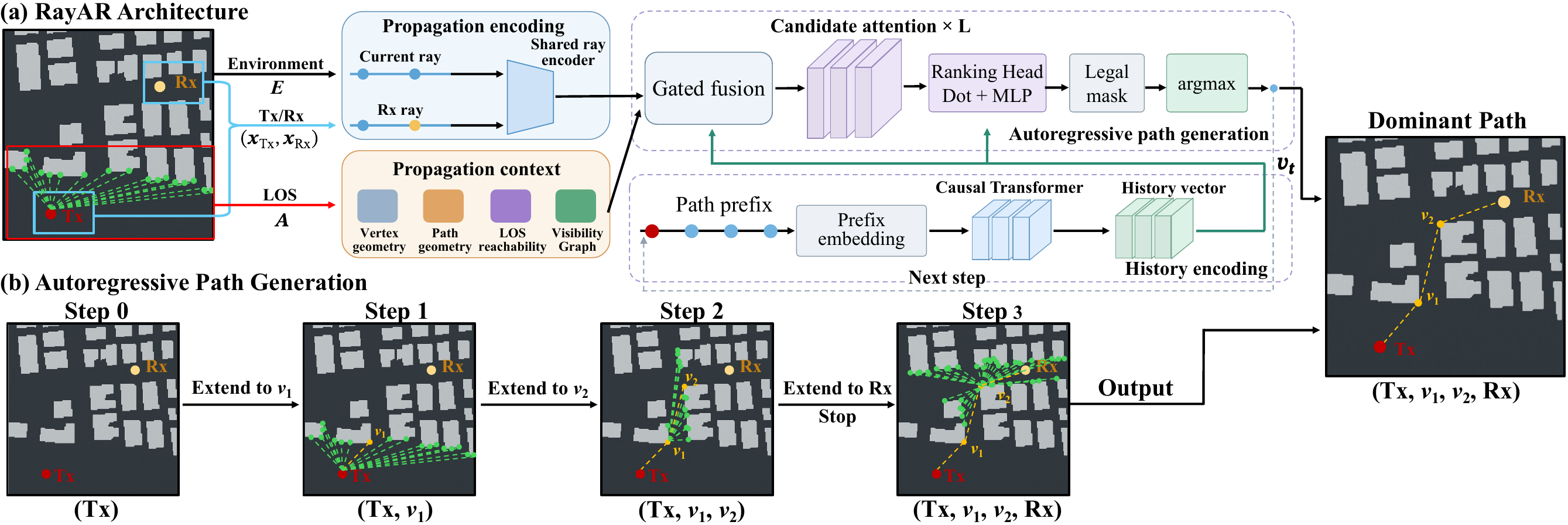}
\caption{Overview of RayAR. (a) Candidate encoding and ranking. (b) Autoregressive path generation.}
\label{fig:framework}
\label{fig:rayar_overview}
\end{figure}

\section{System Model and Problem Formulation}
\label{sec:system}

Tx positions and building geometry are fixed across link queries, while Rx positions vary. Building corners define a candidate set $V$ of diffraction vertices. Building geometry and specified heights determine the 3D vertex coordinates $\bm{x}_v$. The 3D Tx and Rx coordinates are specified separately. The binary relation $A(a,b)$ specifies visibility between locations $a$ and $b$. It covers connections between vertices, from Tx to vertices, and from vertices to Rx. The vertex visibility graph uses $V$ as its node set and connects $u,v\in V$ when $A(u,v)=1$. We denote the query inputs by $X=(E,\bm{x}_{\mathrm{Tx}},\bm{x}_{\mathrm{Rx}},A)$. Here, $E$ is the building geometry, and $\bm{x}_{\mathrm{Tx}}$ and $\bm{x}_{\mathrm{Rx}}$ are the 3D Tx and Rx coordinates. For the studied NLoS links, RayAR selects one to three distinct vertices from $V$ to estimate a dominant path,
\begin{equation}
P=(\mathrm{Tx},v_1,\ldots,v_K,\mathrm{Rx}),
\quad 1\leq K\leq 3,
\label{eq:path}
\end{equation}
where $K$ is the interaction count. Every path segment requires line of sight (LOS) under the supplied visibility relation. The vertex sequence determines the path segments and the incoming and outgoing geometry at each diffraction interaction. D$^2$LoS~\cite{wang2026d2los} combines learning with geometric refinement to accelerate visibility construction from Tx to vertices. For fixed building geometry, the vertex visibility graph can be computed in advance and reused across link queries. RayAR uses these visibility relations as a geometric prior.

We construct ground truth paths offline from the RT paths and path losses stored in RayVerse-100. Let $R$ contain the eligible diffraction paths for a link, and let $L(P)$ denote the path loss of $P$ in dB. We first select the stored path with the lowest path loss as $P_0=\operatorname{argmin}_{P\in R}L(P)$. Ties are resolved by path length, interaction count, and vertex coordinates. In the visibility tree construction used by IRT~\cite{altair_irt}, a path branch is completed by a direct LOS segment to Rx. We retain the path prefix of $P_0$ through the first vertex with LOS to Rx and connect that vertex to Rx. This gives the ground truth path $P^*$, whose vertex sequence provides the training targets and is used to evaluate chain accuracy. Propagation parameters are computed from the geometry of $P^*$.

Let $(X_n,h_n,y_n)$ be a training sample, where $h_n$ is a prefix of $P^*$ and $y_n$ its next vertex. For a batch of $N$ samples, $\theta$ parameterizes the next-vertex distribution $p_\theta$. During generation, $v_{<t}=(v_1,\ldots,v_{t-1})$ and $u_{t-1}$ denotes its endpoint, with $u_0=\mathrm{Tx}$.

\begin{problem}
\label{problem-1}
\leavevmode
\begin{subequations}\label{eq:formulation}
\begin{align}
\min_{\theta}\quad & -\frac{1}{N}\sum_{n=1}^{N}
\log p_\theta(y_n\mid h_n,X_n),
\label{eq:loss}\\
\mathrm{s.t.}\quad & C_t=\{v\in V:A(u_{t-1},v)=1\}
\nonumber\\
&\phantom{C_t={}}\setminus\{v_1,\ldots,v_{t-1}\},
\label{eq:candidates}\\
&p_\theta(v\mid v_{<t},X)=0,\quad v\notin C_t,
\label{eq:legal_support}\\
&v_t=\underset{v\in C_t}{\operatorname{argmax}}\;
 p_\theta(v\mid v_{<t},X),
\label{eq:selection}\\
&K=\min\{t\in\{1,2,3\}:A(v_t,\mathrm{Rx})=1\}.
\label{eq:completion}
\end{align}
\end{subequations}
\end{problem}

The objective in~\eqref{eq:loss} fits the next-vertex distribution to prefixes of the dominant reference path $P^*$. Thus, the RT-derived path losses enter learning through the target vertices, while prediction is conditioned on the query inputs $X$ and the selected prefix. This connects loss-based offline path selection with geometry-conditioned path generation. Equations~\eqref{eq:candidates}--\eqref{eq:selection} enforce visible extensions without revisiting a vertex. Equation~\eqref{eq:completion} gives the first LOS completion within three interactions. The decoding relations apply to active, nonempty candidate sets. An empty $C_t$ while Rx is blocked, or failure to complete within three vertices, yields an incomplete prediction.

For the path in \eqref{eq:path}, set $\bm{x}_0=\bm{x}_{\mathrm{Tx}}$, $\bm{x}_{K+1}=\bm{x}_{\mathrm{Rx}}$, and $\bm{x}_i=\bm{x}_{v_i}$ for intermediate vertices. Propagation delay follows from the total segment length,
\begin{equation}
\tau(P)=\frac{1}{c}\sum_{i=0}^{K}
\|\bm{x}_{i+1}-\bm{x}_i\|_2,
\label{eq:delay}
\end{equation}
where $c$ is the speed of light. The first path segment defines DoD, and the final path segment toward Rx defines DoA. Path loss is evaluated using the uniform theory of diffraction (UTD)~\cite{kouyoumjian1974utd,wang2026d2los}.

\section{RayAR for Link Level Path Estimation}
\label{sec:method}

\subsection{Propagation Feature Encoding}
\label{sec:encoding}

RayAR encodes each legal candidate $v\in C_t$ through the six feature branches in Fig.~\ref{fig:framework}(a). The current ray extends from the current point to $v$ and describes incoming geometry at $v$. The Rx ray describes geometry from $v$ toward Rx, even when that segment is blocked. Both rays use a shared ray encoder with distinct role embeddings. Segment geometry and endpoint attributes form a ray token. Wall tokens are ordered by distance to the ray and describe wall geometry, position, and orientation in its local frame. A Transformer processes these tokens jointly, and its ray token output gives the ray representation.

Vertex geometry describes each candidate and its adjacent walls, while path geometry records distances and directions relative to Tx, the current point, and Rx, together with the current turn, path depth, and accumulated length. LOS reachability features use the vertex visibility graph and vertex-to-Rx visibility to describe whether a candidate can reach Rx within the remaining interaction count, the shortest connection length, and the geometry of its first and final segments. At the initial step, a graph encoder aggregates vertex geometry, visibility to Rx, and connection geometry over the graph to guide the first vertex selection.

\subsection{Autoregressive Path Generation}
\label{sec:generation}

For query inputs $X$, RayAR generates the vertex sequence through the autoregressive factorization
\begin{equation}
p_\theta(v_1,\ldots,v_K\mid X)
=\prod_{t=1}^{K}p_\theta(v_t\mid v_{<t},X).
\label{eq:autoregressive}
\end{equation}
Each conditional distribution is defined over $C_t$ in~\eqref{eq:candidates}. The history encoder summarizes the path prefix through a causal Transformer. Its learned start token combines current path geometry with the Tx/Rx coordinates, followed by selected vertex tokens in path order. Each vertex token combines vertex geometry and a position embedding with physical features encoded by a multilayer perceptron (MLP), including segment and accumulated lengths, distance to Rx, and incoming and outgoing directions along the prefix. The last valid Transformer output gives the history vector, which conditions gated fusion. Each of the six feature branches is projected to a common dimension and split into channel groups, whose gates depend on both projected features and the history vector.

The gated features are concatenated and projected into fused candidate representations. Candidate attention compares them through self-attention and feedforward layers, after which the ranking head combines each representation with the history vector and visibility to Rx. A scaled dot product and an MLP score determine each candidate's score. Masking illegal candidates and applying softmax gives the probabilities in~\eqref{eq:autoregressive}.

The model is trained with ground truth prefixes using~\eqref{eq:loss} and applies the same decoder recursively at inference. Starting from Tx, it appends the most likely legal vertex and updates the current ray, path geometry, LOS reachability features, and history vector before the next decision. Once the current endpoint has LOS to Rx, the path is completed by a direct connection. Fig.~\ref{fig:framework}(b) illustrates this process with two interaction vertices, where green rays indicate legal extensions and yellow segments indicate the selected prefix.

\section{Experiments}
\label{sec:experiments}

\subsection{Experimental Settings}
\label{sec:settings}

We use RayVerse-100 from D$^2$LoS~\cite{wang2026d2los}, with 100 environments and 100 Tx positions per environment. Each environment covers $256\times256$~m with 20~m building heights. Building geometry provides the inputs, and 6~GHz RT records provide ground truth paths and losses. We retain NLoS links with one to three distinct diffraction vertices, excluding rooftop-only paths. Spatially balanced sampling selects 200 Rx positions per Tx. For each Tx, an 80:20 Rx split defines training and validation, stratified by interaction count and Tx--Rx distance with seed 2026. RayAR is trained with AdamW. Baselines include autoregressive (AR) and non-AR methods.

Chain accuracy is the fraction of exact vertex sequence matches. Incomplete predictions count as incorrect. DoD and DoA errors are 3D angular differences. Delay error is the absolute delay difference in ns. For fixed transmit power, power error is
\begin{equation}
\text{Power error}=\max(0,L_{\mathrm{pred}}-L_{\mathrm{GT}}),
\label{eq:loss_metrics}
\end{equation}
where $L_{\mathrm{pred}}$ and $L_{\mathrm{GT}}$ are predicted and ground truth path losses in dB. By the dominant path definition, $L_{\mathrm{pred}}-L_{\mathrm{GT}}$ is nonnegative for feasible predictions. Relative loss and delay errors divide absolute errors by $|L_{\mathrm{GT}}|$ and ground truth delay and are averaged over links. Losses use cached RT values when available for both paths and the same UTD implementation otherwise. All predictions, including incomplete ones, are evaluated from Tx through the selected vertices to Rx.

\begin{figure}[!t]
\centering
\includegraphics[width=\linewidth]{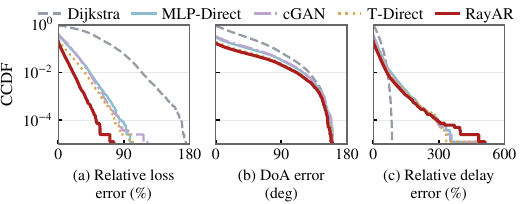}
\caption{CCDFs of propagation errors against non-AR baselines.}
\label{fig:cdf}
\end{figure}

\begin{figure}[!t]
\centering
\includegraphics[width=\linewidth]{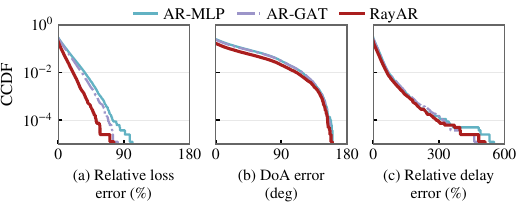}
\caption{CCDFs of propagation errors against AR baselines.}
\label{fig:cdf_ar}
\end{figure}

\subsection{Path Estimation Performance}
\label{sec:performance}
\label{sec:ar_comparison}

Dijkstra~\cite{dijkstra1959} finds the shortest path on the vertex visibility graph. MLP-Direct and Transformer-Direct (T-Direct) predict the complete vertex sequence in one forward pass. A conditional generative adversarial network (cGAN)~\cite{mirza2014cgan} uses adversarial training for complete vertex sequence prediction. In Table~\ref{tab:performance}, RayAR reaches 79.69\% chain accuracy, exceeding the strongest non-AR baseline, T-Direct, by 4.96 points. Mean power error falls from 3.37~dB to 1.76~dB.

\begin{table}[!htb]
\centering
\caption{Accuracy against non-AR methods.}
\label{tab:performance}
\footnotesize
\setlength{\tabcolsep}{2pt}
\begin{tabular*}{\columnwidth}{@{\extracolsep{\fill}}lrrrr>{\columncolor{ourrow}}r}
\toprule
Metric & Dijkstra & MLP-Direct & cGAN & T-Direct & RayAR\\
\midrule
Chain acc. (\%) $\uparrow$ & 5.26 & 61.79 & 57.58 & 74.73 & \textcolor{ourred}{\textbf{79.69}}\\
Power err. (dB) $\downarrow$ & 61.50 & 6.14 & 6.11 & 3.37 & \textcolor{ourred}{\textbf{1.76}}\\
Rel. loss err. (\%) $\downarrow$ & 41.33 & 4.83 & 5.24 & 2.63 & \textcolor{ourred}{\textbf{1.29}}\\
DoD err. ($^\circ$) $\downarrow$ & 33.12 & 9.14 & 9.52 & 6.20 & \textcolor{ourred}{\textbf{5.73}}\\
DoA err. ($^\circ$) $\downarrow$ & 32.68 & 11.48 & 12.62 & 7.74 & \textcolor{ourred}{\textbf{6.72}}\\
Delay err. (ns) $\downarrow$ & 161.21 & 50.75 & 53.52 & 33.36 & \textcolor{ourred}{\textbf{27.37}}\\
Rel. delay err. (\%) $\downarrow$ & 20.77 & 7.57 & 7.76 & 5.10 & \textcolor{ourred}{\textbf{4.10}}\\
\bottomrule
\end{tabular*}
\end{table}

AR-MLP and AR-GAT~\cite{velickovic2018gat} use an MLP and graph attention, respectively, to select vertices sequentially and update inputs after each selection. Table~\ref{tab:performance_ar} shows that RayAR improves chain accuracy by 9.69 points over AR-MLP and 8.03 points over AR-GAT.

\begin{table}[!htb]
\centering
\caption{Accuracy among AR methods.}
\label{tab:performance_ar}
\footnotesize
\setlength{\tabcolsep}{2pt}
\begin{tabular*}{\columnwidth}{@{\extracolsep{\fill}}lrr>{\columncolor{ourrow}}r}
\toprule
Metric & AR-MLP & AR-GAT & RayAR\\
\midrule
Chain acc. (\%) $\uparrow$ & 70.00 & 71.66 & \textcolor{ourred}{\textbf{79.69}}\\
Power err. (dB) $\downarrow$ & 4.46 & 3.84 & \textcolor{ourred}{\textbf{1.76}}\\
Rel. loss err. (\%) $\downarrow$ & 3.03 & 2.62 & \textcolor{ourred}{\textbf{1.29}}\\
DoD err. ($^\circ$) $\downarrow$ & 9.32 & 9.01 & \textcolor{ourred}{\textbf{5.73}}\\
DoA err. ($^\circ$) $\downarrow$ & 10.14 & 9.70 & \textcolor{ourred}{\textbf{6.72}}\\
Delay err. (ns) $\downarrow$ & 40.68 & 39.70 & \textcolor{ourred}{\textbf{27.37}}\\
Rel. delay err. (\%) $\downarrow$ & 5.99 & 5.80 & \textcolor{ourred}{\textbf{4.10}}\\
\bottomrule
\end{tabular*}
\end{table}

\label{sec:errors}

Figs.~\ref{fig:cdf} and~\ref{fig:cdf_ar} show complementary cumulative distribution functions (CCDFs), with Tables~\ref{tab:percentiles} and~\ref{tab:percentiles_ar} reporting the 90th and 95th error percentiles (P90 and P95). At P95, RayAR reduces relative loss error from T-Direct's 17.77\% to 8.92\% and relative delay error from 31.87\% to 27.46\%.

\begin{table}[!htb]
\centering
\caption{Error percentiles against non-AR methods.}
\label{tab:percentiles}
\footnotesize
\setlength{\tabcolsep}{2pt}
\begin{tabular*}{\columnwidth}{@{\extracolsep{\fill}}lrrrr>{\columncolor{ourrow}}r}
\toprule
Metric & Dijkstra & MLP-Direct & cGAN & T-Direct & RayAR\\
\midrule
Rel. loss err. P90 (\%) $\downarrow$ & 80.14 & 17.66 & 18.63 & 9.14 & \textcolor{ourred}{\textbf{4.16}}\\
Rel. loss err. P95 (\%) $\downarrow$ & 92.95 & 26.37 & 26.43 & 17.77 & \textcolor{ourred}{\textbf{8.92}}\\
Rel. delay err. P90 (\%) $\downarrow$ & 44.11 & 25.61 & 25.94 & 16.84 & \textcolor{ourred}{\textbf{12.86}}\\
Rel. delay err. P95 (\%) $\downarrow$ & 52.44 & 40.25 & 39.44 & 31.87 & \textcolor{ourred}{\textbf{27.46}}\\
\bottomrule
\end{tabular*}
\end{table}

\begin{table}[!htb]
\centering
\caption{Error percentiles among AR methods.}
\label{tab:percentiles_ar}
\footnotesize
\setlength{\tabcolsep}{2pt}
\begin{tabular*}{\columnwidth}{@{\extracolsep{\fill}}lrr>{\columncolor{ourrow}}r}
\toprule
Metric & AR-MLP & AR-GAT & RayAR\\
\midrule
Rel. loss err. P90 (\%) $\downarrow$ & 11.07 & 9.54 & \textcolor{ourred}{\textbf{4.16}}\\
Rel. loss err. P95 (\%) $\downarrow$ & 18.53 & 16.40 & \textcolor{ourred}{\textbf{8.92}}\\
Rel. delay err. P90 (\%) $\downarrow$ & 21.31 & 20.64 & \textcolor{ourred}{\textbf{12.86}}\\
Rel. delay err. P95 (\%) $\downarrow$ & 35.19 & 34.82 & \textcolor{ourred}{\textbf{27.46}}\\
\bottomrule
\end{tabular*}
\end{table}

\label{sec:hops}

\begin{figure}[!htb]
\centering
\includegraphics[width=\linewidth]{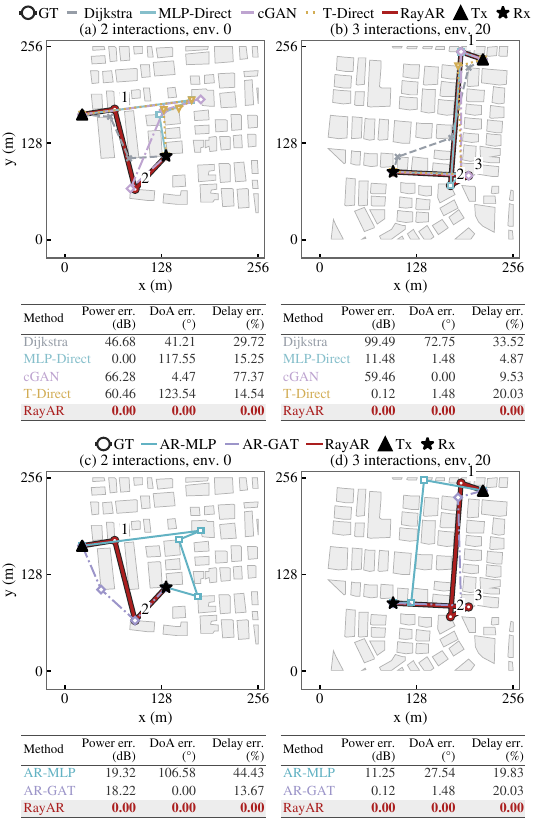}
\caption{Path examples against non-AR baselines in (a)--(b) and AR baselines in (c)--(d).}
\label{fig:cases}
\label{fig:cases_ar}
\end{figure}

Fig.~\ref{fig:cases} compares two- and three-interaction paths against non-AR and AR baselines. AR-GAT in Fig.~\ref{fig:cases}(c) and cGAN in Fig.~\ref{fig:cases}(b) recover the final segment but select different earlier vertices. Their DoA errors are zero, while power and delay errors remain. MLP-Direct in Fig.~\ref{fig:cases}(b) selects a different final diffraction vertex. RayAR matches both ground truth paths. These examples show why arrival direction alone does not identify the full path.

\begin{figure}[!htbp]
\centering
\includegraphics[width=\linewidth]{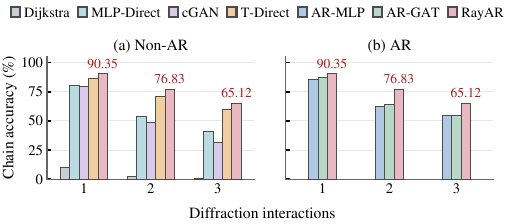}
\caption{Chain accuracy by hop count for non-AR and AR methods.}
\label{fig:hops}
\label{fig:hops_ar}
\end{figure}

Fig.~\ref{fig:hops} reports chain accuracy by hop count, defined as the number of ground truth diffraction interactions. Accuracy decreases with hop count for all methods shown. RayAR leads at all three hop counts, exceeding T-Direct by 6.19 points for two hops and 5.25 points for three hops.

\subsection{Ablation Study}
\label{sec:ablation}

\begin{figure}[!ht]
\centering
\includegraphics[width=\linewidth]{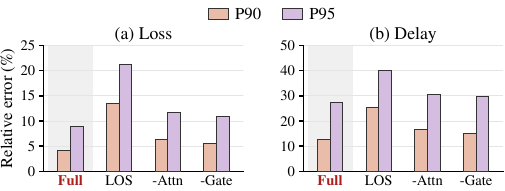}
\caption{Relative error percentiles for Full and ablated RayAR.}
\label{fig:ablation}
\end{figure}

Table~\ref{tab:ablation} and Fig.~\ref{fig:ablation} test LOS features, candidate attention, and gated fusion. The LOS-mask-only variant retains the visibility constraints and completion rule. It removes LOS reachability features, candidate visibility to Rx from the ranking head, and visibility graph features. This causes the largest accuracy drop, from 79.69\% to 65.34\%, showing the value of LOS features for candidate ranking. The candidate attention ablation removes both self-attention and feedforward layers. Replacing gated fusion with feature concatenation retains the history encoder and ranking head. Both changes lower chain accuracy. All three variants increase relative loss and delay errors at P90 and P95.

\begin{table}[!htb]
\centering
\caption{Ablation results for RayAR.}
\label{tab:ablation}
\footnotesize
\setlength{\tabcolsep}{2pt}
\begin{tabular}{lrrrr}
\toprule
Variant & Chain acc. & Power err. & Rel. loss err. & Rel. delay err.\\
 & (\%) $\uparrow$ & (dB) $\downarrow$ & (\%) $\downarrow$ & (\%) $\downarrow$\\
\midrule
\rowcolor{ourrow}
Full RayAR & \textcolor{ourred}{\textbf{79.69}} & \textcolor{ourred}{\textbf{1.76}} & \textcolor{ourred}{\textbf{1.29}} & \textcolor{ourred}{\textbf{4.10}}\\
LOS mask only & 65.34 & 5.16 & 3.70 & 7.47\\
w/o candidate attention & 74.87 & 2.51 & 1.79 & 4.78\\
w/o gated fusion & 77.11 & 2.23 & 1.62 & 4.59\\
\bottomrule
\end{tabular}
\end{table}

\subsection{Inference Efficiency}
\label{sec:efficiency}

RayAR averages 28.67~ms per link in Table~\ref{tab:efficiency}. All RayAR timing results use the same RTX PRO 6000 GPU and batch size 128. Timing includes feature construction, GPU transfer, and greedy decoding, averaged per link over the batch. The IRT time of 24.9~min/scenario reported in D$^2$LoS includes visibility construction, path search, and UTD computation.

\begin{table}[!htb]
\centering
\caption{Computation time.}
\label{tab:efficiency}
\small
\setlength{\tabcolsep}{3pt}
\begin{tabular*}{\columnwidth}{@{\extracolsep{\fill}}lcccc}
\toprule
Type & \multicolumn{4}{c}{Method}\\
\midrule
RT & IRT~\cite{wang2026d2los} & \multicolumn{3}{c}{24.9~(min/scenario)}\\
\midrule
Non-AR & Dijkstra & MLP-Direct & cGAN & T-Direct\\
Time (ms/link) & 9.03 & 132.56 & 131.66 & 182.88\\
\midrule
AR & AR-MLP & \multicolumn{2}{c}{AR-GAT} & RayAR\\
Time (ms/link) & 24.43 & \multicolumn{2}{c}{27.96} & 28.67\\
\bottomrule
\end{tabular*}
\end{table}

Table~\ref{tab:acceleration} evaluates geometric feature caching and bfloat16 (BF16) on 400,000 queries after cache loading and warmup. BF16 with caching averages around 4~ms per link with 79.70\% chain accuracy, a $6.93\times$ speedup over single precision (FP32). Each map's cache takes 3.79~min to build and 1.08~GiB to store on average, and is reused across queries.

\begin{table}[!htb]
\centering
\caption{RayAR inference acceleration.}
\label{tab:acceleration}
\normalsize
\setlength{\tabcolsep}{3pt}
\begin{tabularx}{\columnwidth}{>{\raggedright\arraybackslash}Xrrr}
\toprule
Setting & Chain acc. (\%) & Time (ms/link) & Speedup\\
\midrule
FP32, Table~\ref{tab:efficiency} & 79.69 & 28.67 & 1.00$\times$\\
BF16 & 79.70 & 16.51 & 1.74$\times$\\
FP32 + cache & 79.69 & 7.50 & 3.82$\times$\\
\rowcolor{ourrow}
BF16 + cache & \textcolor{ourred}{\textbf{79.70}} & \textcolor{ourred}{\textbf{4.14}} & \textcolor{ourred}{\textbf{6.93}$\boldsymbol{\times}$}\\
\bottomrule
\end{tabularx}
\end{table}

\subsection{Beam Alignment}
\label{sec:beam}

3GPP Release~18 TR~38.843 studies beam prediction for 5G-Advanced~\cite{3gpp38843}. BM-Case1 predicts beams in Set A from measurements in Set B. We use its evaluation framework for geometry-guided probing on 400,000 validation links. Set A combines 32 Tx discrete Fourier transform (DFT) beams from a $4\times8$ array and eight Rx beams from two $1\times4$ panels, with half-wavelength spacing. We reuse the 6~GHz RayVerse-100 path geometry and recompute each path's UTD loss $L_p$ and diffraction phase $\phi_p$ at 30~GHz. Path length gives delay $\tau_p$. The complex gain of path $p$ is $10^{-L_p/20}e^{j(\phi_p-2\pi f\tau_p)}$ at subcarrier frequency $f$. We coherently sum the beam-weighted paths and average power over 792 subcarriers spaced 120~kHz apart.

We rank pairs by array responses at predicted DoD and DoA, probe the top $B\in\{1,2,4,8\}$ pairs, and select the pair with the highest layer 1 reference signal received power (L1-RSRP). Fixed-Pattern Sweep probes the first $B$ pairs from the zero-based sequence $(10,1)$, $(30,7)$, $(14,3)$, $(26,5)$, $(8,0)$, $(12,2)$, $(24,4)$, $(28,6)$. Rx orientation is fixed and measurements are ideal. Hit rate measures recovery of the full scan optimum. Mean L1-RSRP gap measures power loss from that optimum.

\begin{table}[!htb]
\centering
\caption{Beam alignment across probing budgets.}
\label{tab:beam}
\small
\setlength{\tabcolsep}{2pt}
\renewcommand{\arraystretch}{1.13}
\begin{tabular*}{\columnwidth}{@{\extracolsep{\fill}}l r@{\,/\,}r r@{\,/\,}r r@{\,/\,}r r@{\,/\,}r}
\toprule
Method & \multicolumn{8}{c}{Hit rate (\%) / Mean L1-RSRP gap (dB)}\\
\cmidrule(l){2-9}
 & \multicolumn{2}{c}{$B=1$} & \multicolumn{2}{c}{$B=2$} & \multicolumn{2}{c}{$B=4$} & \multicolumn{2}{c}{$B=8$}\\
\midrule
\shortstack[l]{Fixed-Pattern\\Sweep} & 0.55 & 37.63 & 0.55 & 32.40 & 1.36 & 26.62 & 3.11 & 18.20\\
\midrule
Dijkstra & 19.61 & 17.46 & 25.84 & 14.64 & 32.20 & 12.15 & 40.08 & 9.40\\
MLP-Direct & 63.35 & 4.58 & 68.83 & 3.91 & 72.71 & 3.32 & 76.69 & 2.64\\
cGAN & 61.96 & 4.90 & 67.58 & 4.19 & 71.58 & 3.54 & 75.83 & 2.81\\
T-Direct & 69.34 & 3.20 & 74.48 & 2.77 & 77.67 & 2.39 & 80.81 & 1.96\\
\midrule
AR-MLP & 66.27 & 4.13 & 71.54 & 3.56 & 74.83 & 3.07 & 78.17 & 2.52\\
AR-GAT & 67.33 & 3.93 & 72.45 & 3.39 & 75.68 & 2.93 & 78.87 & 2.40\\
\rowcolor{ourrow}
\textbf{RayAR} & \textcolor{ourred}{\textbf{71.24}} & \textcolor{ourred}{\textbf{2.85}} & \textcolor{ourred}{\textbf{76.14}} & \textcolor{ourred}{\textbf{2.49}} & \textcolor{ourred}{\textbf{79.11}} & \textcolor{ourred}{\textbf{2.16}} & \textcolor{ourred}{\textbf{81.89}} & \textcolor{ourred}{\textbf{1.81}}\\
\bottomrule
\end{tabular*}
\end{table}

At $B=4$, RayAR outperforms AR-GAT at $B=8$ in both metrics in Table~\ref{tab:beam}, halving the probing budget. At $B=8$, RayAR improves hit rate by 1.08 points and mean L1-RSRP gap by 0.15~dB over T-Direct. It places 83.41\% of links within 1~dB of the full scan optimum using 96.88\% fewer probes.

Across all budgets, RayAR achieves the highest hit rate and lowest L1-RSRP gap. At $B=1$, its 1.90-point hit-rate gain and 0.35-dB gap reduction over T-Direct show the value of path-guided probing under a tight budget.

\subsection{Real-world Measurement Experiment}
\label{sec:measurements}

\begin{figure}[!ht]
\centering
\includegraphics[width=\linewidth,pagebox=cropbox]{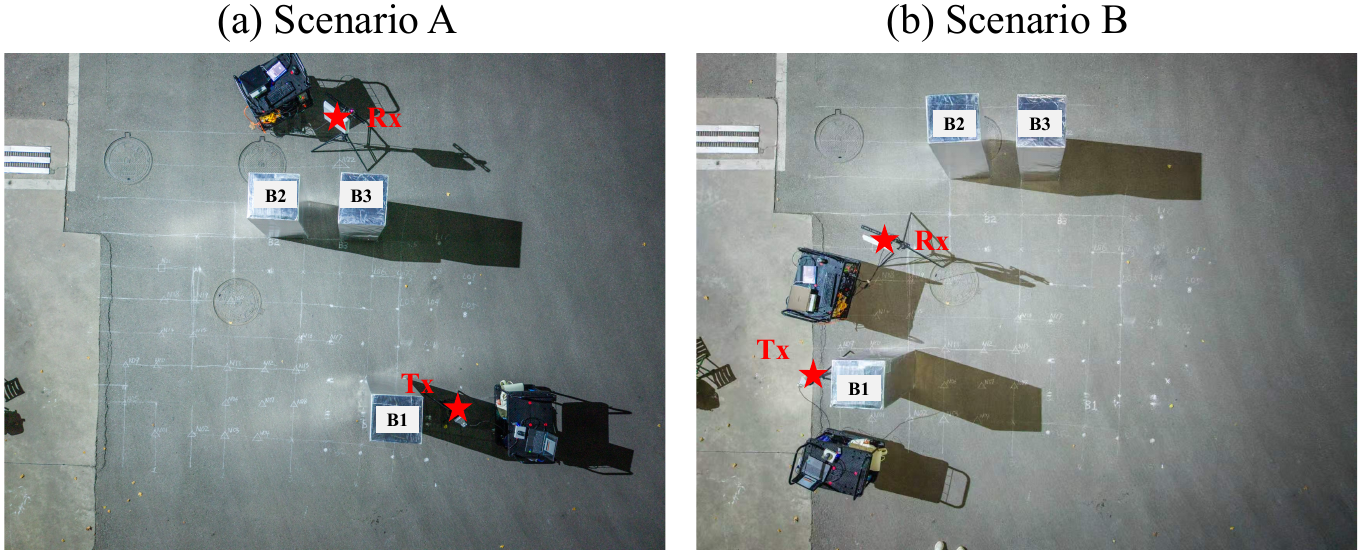}
\caption{Measurement scenarios with Rx positions marked on the ground.}
\label{fig:measurement_scenarios}
\end{figure}

Measurements cover 62 points in the two outdoor layouts in Fig.~\ref{fig:measurement_scenarios}, sized $5.5\times5.0$~m and $6.0\times5.0$~m. Each has a fixed Tx and three boxes about 1~m high wrapped in aluminum foil. The directional Tx and panel Rx antennas have equal heights and vertical polarization. RayAR uses simulated-path weights without fine-tuning. Angular scans give the maximum received signal strength (RSS) and its DoD and DoA ground truth. Predicted RSS uses 6~GHz free space path loss over the predicted 2D path length. A shared offset is calibrated on the 20 LOS links as the median of measured RSS plus modeled path loss. Relative RSS error normalizes absolute error by the magnitude of measured RSS in dBm.

Completion requires LOS for every segment of the generated Tx-to-Rx path and no repeated vertices. The 20 LOS links use direct routing. Table~\ref{tab:measurements} reports completion for all 62 points and the 42 NLoS points. Mean absolute error (MAE) includes all 62 predictions. RayAR, AR-GAT, and AR-MLP complete all NLoS links. cGAN completes only four and frequently produces blocked segments. RayAR achieves the lowest DoA and DoD MAEs, with calibrated RSS MAE close to AR-GAT.

RayAR and AR-GAT have nearly identical calibrated RSS errors because the RSS model mainly depends on path length. RayAR retains lower angular errors and complete NLoS paths through stepwise visibility constraints.

\begin{table}[!htb]
\centering
\caption{Measured link results for non-AR and AR methods.}
\label{tab:measurements}
\label{tab:measurement_validity}
\normalsize
\setlength{\tabcolsep}{2pt}
\renewcommand{\arraystretch}{1.18}
\begin{tabularx}{\columnwidth}{>{\raggedright\arraybackslash}Xlccccc}
\toprule
Type & Method &
\shortstack{Completion\\(\%) $\uparrow$\\All/NLoS} &
\shortstack{RSS\\MAE\\(dB) $\downarrow$} &
\shortstack{RSS\\Rel.\\(\%) $\downarrow$} &
\shortstack{DoA\\MAE\\($^\circ$) $\downarrow$} &
\shortstack{DoD\\MAE\\($^\circ$) $\downarrow$}\\
\midrule
\multirow{3}{*}{\rotatebox[origin=c]{90}{Non-AR}}
& \shortstack[l]{MLP-\\Direct} & 74.2/61.9 & 5.10 & 7.71 & 21.43 & 25.93\\
& cGAN & 38.7/9.5 & 4.50 & 6.83 & 24.53 & 22.29\\
& T-Direct & 50.0/26.2 & 5.32 & 8.04 & 27.54 & 24.32\\
\midrule
\multirow{3}{*}{\rotatebox[origin=c]{90}{AR}}
& AR-MLP & \textbf{100.0/100.0} & 5.14 & 7.76 & 21.32 & 15.24\\
& AR-GAT & \textbf{100.0/100.0} & \textbf{3.89} & \textbf{5.88} & 9.50 & 11.53\\
\rowcolor{ourrow}
& \textbf{RayAR} & \textbf{100.0/100.0} & 3.90 & 5.90 & \textbf{9.07} & \textbf{11.16}\\
\bottomrule
\end{tabularx}
\end{table}

\section{Conclusion}
\label{sec:conclusion}

RayAR uses autoregressive vertex selection for fast dominant diffraction path estimation. On RayVerse-100, it achieves 79.69\% chain accuracy and 1.76~dB mean power error. Geometric feature caching and BF16 reduce inference to around 4~ms per link. Without fine-tuning, RayAR completes all measured NLoS links and achieves the lowest DoD and DoA errors among the compared methods. Its paths also reduce the beam alignment probing budget.


\bibliographystyle{IEEEtran}
\bibliography{refs}

@techreport{3gpp38843,
  author = {{3GPP}},
  title = {Study on Artificial Intelligence ({AI})/Machine Learning ({ML}) for {NR} Air Interface},
  institution = {{3GPP}},
  type = {TR},
  number = {38.843 V18.0.0},
  year = {2023}
}

@article{wang2025radiodiff,
  author = {Xiucheng Wang and Keda Tao and Nan Cheng and Zhisheng Yin and Zan Li and Yuan Zhang and Xuemin Shen},
  title = {{RadioDiff}: An Effective Generative Diffusion Model for Sampling-Free Dynamic Radio Map Construction},
  journal = {IEEE Trans. Cogn. Commun. Netw.},
  volume = {11},
  number = {2},
  pages = {738--750},
  year = {2025},
  doi = {10.1109/TCCN.2024.3504489}
}

@misc{huang2026map2aps,
  author = {Junxi Huang and Xiucheng Wang and Nan Cheng and Kailong Wang and Ruijin Sun and Zhisheng Yin},
  title = {{Map2APS}: A Physically Grounded Benchmark for Direct Angle Power Spectrum Prediction from Urban Geometry},
  howpublished = {arXiv:2605.14989},
  year = {2026}
}

@article{kim2024raytracing,
  author = {Yongwan Kim and Hyunjun Yang and Hooyoung Kim and Junpyo Jo and Jungsuek Oh},
  title = {Anxel Beam Shrinkage Method and Heterogeneous Computing-Accelerated Full-Image Theory Method Ray Tracing Enabling Massive Outdoor Propagation Modeling},
  journal = {IEEE Trans. Antennas Propag.},
  volume = {72},
  number = {7},
  pages = {5935--5949},
  year = {2024},
  doi = {10.1109/TAP.2024.3411793}
}

@inproceedings{wahl2005dominant,
  author = {Ren{\'e} Wahl and Gerd W{\"o}lfle and Philipp Wertz and Pascal Wildbolz and Friedrich Landstorfer},
  title = {Dominant Path Prediction Model for Urban Scenarios},
  booktitle = {Proc. 14th IST Mobile Wireless Commun. Summit},
  year = {2005}
}

@misc{wang2026d2los,
  author = {Xiucheng Wang and Junxi Huang and Conghao Zhou and Xuemin Shen and Nan Cheng},
  title = {Physics-Informed Line-of-Sight Learning for Scalable Deterministic Channel Modeling},
  howpublished = {arXiv:2603.27976},
  year = {2026},
  doi = {10.48550/arXiv.2603.27976}
}

@inproceedings{jin2025sandwich,
  author = {Yifei Jin and Ali Maatouk and Sarunas Girdzijauskas and Shugong Xu and Leandros Tassiulas and Rex Ying},
  title = {{SANDWICH}: Towards an Offline, Differentiable, Fully-Trainable Wireless Neural Ray-Tracing Surrogate},
  booktitle = {Proc. IEEE Int. Conf. Mach. Learn. Commun. Netw. (ICMLCN)},
  pages = {1--7},
  year = {2025},
  doi = {10.1109/ICMLCN64995.2025.11139897}
}

@inproceedings{eertmans2025sampling,
  author = {J{\'e}rome Eertmans and Nicola {Di Cicco} and Claude Oestges and Laurent Jacques and Enrico Maria Vitucci and Vittorio Degli-Esposti},
  title = {Towards Generative Ray Path Sampling for Faster Point-to-Point Ray Tracing},
  booktitle = {Proc. IEEE Int. Conf. Mach. Learn. Commun. Netw. (ICMLCN)},
  pages = {1--6},
  year = {2025},
  doi = {10.1109/ICMLCN64995.2025.11140249}
}

@article{kouyoumjian1974utd,
  author = {R. G. Kouyoumjian and P. H. Pathak},
  title = {A Uniform Geometrical Theory of Diffraction for an Edge in a Perfectly Conducting Surface},
  journal = {Proc. IEEE},
  volume = {62},
  number = {11},
  pages = {1448--1461},
  year = {1974},
  doi = {10.1109/PROC.1974.9651}
}

@manual{altair_irt,
  author = {{Altair Engineering Inc.}},
  title = {Intelligent Ray Tracing (IRT)},
  year = {2026},
  note = {{WinProp} Documentation, accessed: Apr. 13, 2026},
  url = {https://help.altair.com/winprop/topics/winprop/user_guide/propagation_methods/propgation_models_indoor_irt.htm}
}

@article{heng2022probing,
  author = {Yuqiang Heng and Jianhua Mo and Jeffrey G. Andrews},
  title = {Learning Site-Specific Probing Beams for Fast mmWave Beam Alignment},
  journal = {IEEE Trans. Wireless Commun.},
  volume = {21},
  number = {8},
  pages = {5785--5800},
  year = {2022},
  doi = {10.1109/TWC.2022.3143121}
}

@article{dijkstra1959,
  author = {E. W. Dijkstra},
  title = {A Note on Two Problems in Connexion with Graphs},
  journal = {Numer. Math.},
  volume = {1},
  number = {1},
  pages = {269--271},
  year = {1959},
  doi = {10.1007/BF01386390}
}

@misc{mirza2014cgan,
  author = {Mehdi Mirza and Simon Osindero},
  title = {Conditional Generative Adversarial Nets},
  howpublished = {arXiv:1411.1784},
  year = {2014}
}

@inproceedings{velickovic2018gat,
  author = {Petar Veli{\v{c}}kovi{\'c} and Guillem Cucurull and Arantxa Casanova and Adriana Romero and Pietro Li{\`o} and Yoshua Bengio},
  title = {Graph Attention Networks},
  booktitle = {Proc. Int. Conf. Learn. Represent. (ICLR)},
  year = {2018}
}

\end{document}